\documentclass{article}

\usepackage{blindtext} 

\usepackage[sc]{mathpazo} 
\usepackage[T1]{fontenc} 
\usepackage{microtype} 

\usepackage[english]{babel} 

\usepackage[hmarginratio=1:1,top=32mm,columnsep=20pt]{geometry} 
\usepackage[hang, small,labelfont=bf,up,textfont=it,up]{caption} 
\usepackage{booktabs} 

\usepackage{lettrine} 

\usepackage{titlesec}
\usepackage{enumitem} 
\setlist[itemize]{noitemsep} 

\usepackage{abstract} 

\usepackage{fancyhdr} 
\usepackage{titling} 
\usepackage{authblk}
\usepackage{verbatim}

\usepackage{hyperref} 

\usepackage{graphicx}
\usepackage{subcaption}
\usepackage{amsmath}

\pretitle{\begin{center}\Huge\bfseries} 
\posttitle{\end{center}} 
\title{Deep Learning for Spectral Filling in Radio Frequency Applications} 
\author[1,*]{Matthew Setzler}
\author{Lizzy Coda}
\author{Jeremiah Rounds}
\author{Michael Vann}
\author{Michael Girard}
\affil{Pacific Northwest National Laboratory}
\affil[*]{Corresponding author: Matthew Setzler, matt.setzler@pnnl.gov}
\date{\today} 

\begin{document}

\maketitle


\begin{abstract}
Due to the Internet of Things (IoT) proliferation, Radio Frequency (RF) channels are increasingly congested with new kinds of devices, which carry unique and diverse communication needs. This poses complex challenges in modern digital communications, and calls for the development of technological innovations that (i) optimize capacity (bitrate) in limited bandwidth environments, (ii) integrate cooperatively with already-deployed RF protocols, and (iii) are adaptive to the ever-changing demands in modern digital communications. In this paper we present methods for applying deep neural networks for \textit{spectral filling}. Given an RF channel transmitting digital messages with a pre-established modulation scheme, we automatically learn novel modulation schemes for sending extra information, in the form of additional messages, ``around'' the fixed-modulation signals (i.e., without interfering with them). In so doing, we effectively increase channel capacity without increasing bandwidth. We further demonstrate the ability to generate signals that closely resemble the original modulations, such that the presence of extra messages is undetectable to third-party listeners. We present three computational experiments demonstrating the efficacy of our methods, and conclude by discussing the implications of our results for modern RF applications.
\end{abstract}

\section{Introduction}
\label{sec:introduction}

The Internet of Things (IoT) proliferation poses novel, and complex challenges for digital communications \cite{miorandi2012internet, bandyopadhyay2011internet}. Radio Frequency (RF) channels are increasingly congested with new kinds of devices, which carry unique communication needs \cite{mukhopadhyay2014internet}. Meeting these challenges requires the development of new technologies that (i) optimize capacity in limited bandwidth environments, (ii) integrate seamlessly with existing, already-deployed communications protocols, and (iii) are adaptive to the continuous flux in consumption requirements of modern digital comms environments.

Here we present novel methods for applying deep neural networks (DNNs) for \textit{spectral filling}. Given an RF channel transmitting digital messages via some pre-established modulation scheme\footnote{The experiments reported in this paper utilitize Binary Phase Shift Keying (BPSK) and Quadrature Phase Shift Keying (QPSK), but in principle this could be any pre-defined modulation scheme.}, we show that we can automatically learn novel modulation schemes to send extra information, in the form of an additional message, ``around'' the fixed-modulation signals (i.e., without interfering with them), thus increasing channel capacity without increasing bandwidth. We further demonstrate the ability to constrain the spectral shape of learned signals, such that they resemble the original modulations or conform to arbitrary spectral shapes.

Recent years have seen a nascent, but growing interest in leveraging deep learning for RF applications. One such application is ``spectrum sensing'', where DNNs are trained to classify the modulations of signals in an RF environment \cite{rajendran2018deep, chandhok2019lstm}. Neural networks have also been trained to demodulate RF signals \cite{ohnishi1996neural, wang2019deep, amini2010improving, lerkvaranyu2004m, onder2013neural, lyu2018performance}, and even for end-to-end communications systems, although success of these efforts has been mixed \cite{dorner2017deep, o2017introduction}. Despite these early efforts, deep learning in RF applications is still a relatively unexplored area, and much remains to be learned about what kinds of model architectures are well-suited to the RF domain and what kinds of problems DNNs are apt to address. 

In particular, the spectrum filling problem introduced in this paper has not yet been addressed by the research community. Earlier efforts have shown that DNNs can be used in model communications systems, but it is not clear how they would be deployed in real-world scenarios, in which the learned RF signals would need to cooperate with existing signals defined by pre-established modulation protocols. Conversely, in the present work, insofar as we are able to learn modulations that adapt to existing RF protocols, we demonstrate the suitability of our methods to be integrated with already-deployed communications systems in the wild.

Our work also differs from previous efforts in that our DNN architectures utilize Transformer networks, which have proven to be powerful architectures for modeling temporal relationships in time-series data such as NLP, music, and signal processing \cite{vaswani2017attention, devlin2018bert, huang2018music, dong2018speech}. This is a departure from previous efforts, which have typically used convolutional networks or residual networks \cite{west2017deep, o2018over}, which were originally developed in computer vision \cite{krizhevsky2012imagenet, he2016deep}, and thus not optimally-suited for modeling time-series. There has been some work on applying autoregressive Long-Short Term Memory (LSTM) networks to RF data \cite{rajendran2018deep, chandhok2019lstm}, but these efforts lag behind the state-of-the-art in deep learning, because LSTMs are almost unanimously outperformed by Transformers in a variety of time-series applications \cite{giuliari2021transformer, vaswani2017attention}. To our knowledge, there has only been one previous application of Transformer networks to the RF domain \cite{shevitski2021digital}, which showed promising results, though it was not geared towards the problem addressed here: spectral filling.

\subsection{Problem Statement: Spectral Filling}

\begin{figure}
\centering
    \includegraphics[width=.7\columnwidth,]{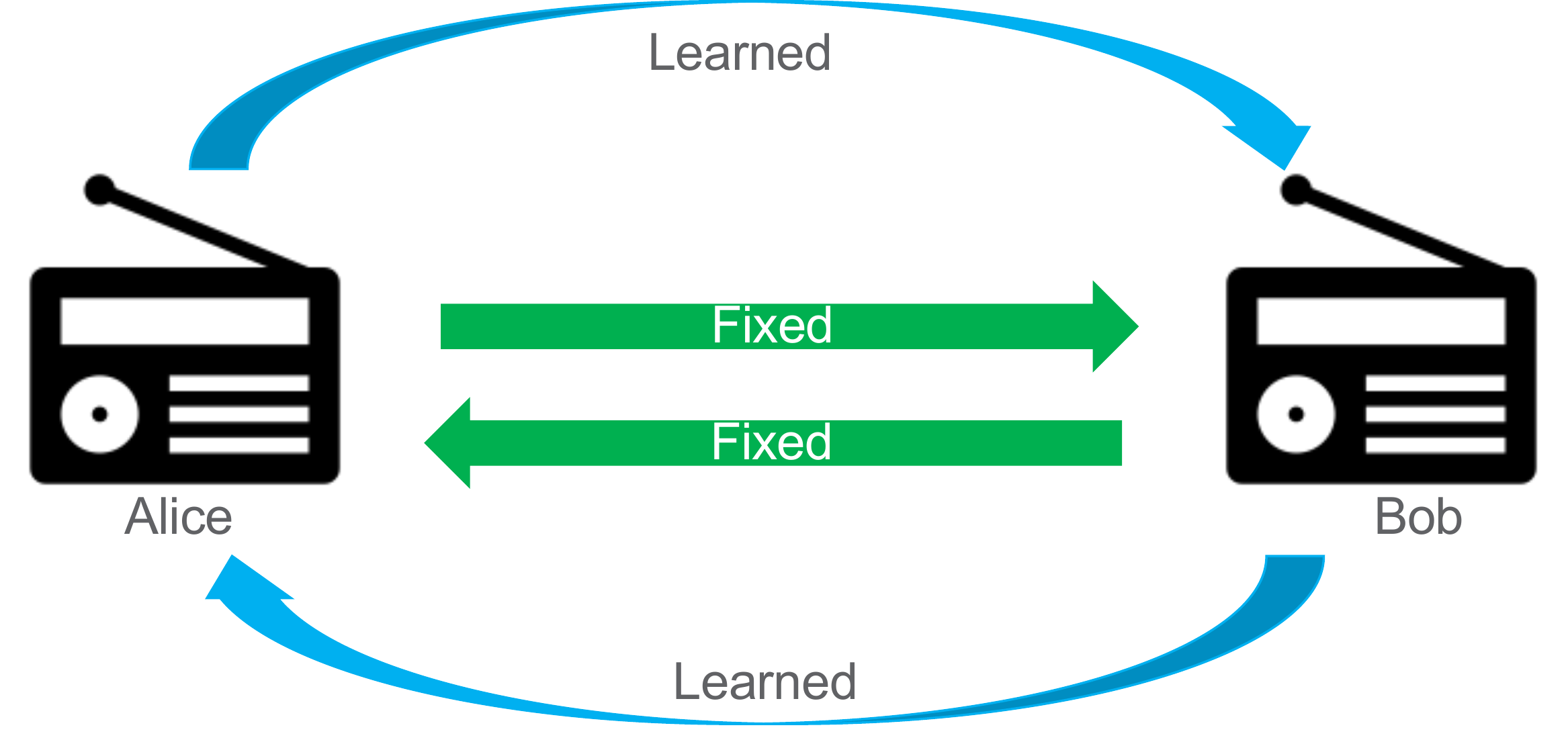}
   \caption{Alice and Bob communicate with a fixed, traditional scheme and fill the bandwidth capacity with the learned communication scheme.}
   \label{fig:Alice-Bob}
\end{figure}

We considered a scenario where two radios communicate over a traditional digital signal pipeline as in Figure~\ref{fig:Alice-Bob}. This communication scheme is bounded in its capacity by Shannon’s Limit ~\cite{shannon1948mathematical}, meaning that for the bandwidth the radios are using and the amount of noise in their environment, the speed that they can transmit information is fixed. This is defined by the equation
\begin{equation}
C= B \log_{2} \left(1+\frac{S}{N}\right)
\end{equation}
where $C$ is the capacity in bits/sec, $B$ is the bandwidth in Hz, $S$ and $N$ are the power in the signal and noise respectively. Modern digital communication schemes can come close to this limit, however there is usually a gap in the actual speed of data transmission and the theoretical maximum. This means that there is the possibility for extra data to be transmitted alongside the fixed, traditional scheme. 

However, another important theorem of digital communications that stops full utilization of this gap is the central coding limit theorem. The coding limit theorem states that while the rate, $R$ of data transfer in a channel is less than Shannon's capacity, $R <C$, the rate at which errors occur in the communication channel can be made arbitrarily small. If the rate exceeds the Shannon limit then the error rate will be, in general, large. We plan to exploit this gap in actual vs. theoretical rates of communication, while still being able to make the error rates of communication small.

We label a traditional digital communication signal from one radio to the other as the A message. This consists of a sequence of ones and zeros and is generally long. If this signal does not reach Shannon’s limit than there is the possibility for a second message that uses some of the unused bandwidth. This is the B message but is generally not as long as the A message. We have developed a novel method for generating a time series that can transmit these two different types of messages without greatly affecting the accuracy of the A message. 

A secondary goal of ours is to constraint properties of learned signals using auxiliary loss terms. In Experiment 2, we constrain learned signals to resemble the original modulations, such that a third-party would not be able to identify the presence of message B based on spectral properties or other signatures of the generated signals. In Experiment 3 we go one step further and show that it is possible to constrain learned signals to match to arbitrary spectral shapes, while still retaining the ability to transmit both messages. In the remainder of this paper we specify our methods, report results from three experiments demonstrating success with respect to each of our goals, and conclude by discussing the implications for modern RF applications.

\section{Methods}
\label{sec:methods}

Our goal is to transmit an RF signal\footnote{RF signals typically comprise two orthogonal components, I and Q, which can be thought of as cosine and sine components of a complex waveform. Sampling from these components yields a two-dimensional IQ sequence, which for the purposes of this paper is synonymous with an RF signal.} that carries information from two messages (A and B) over-the-air. Both messages are sequences of discrete symbols. In Experiment 1, which utilizes Binary Phase Shift Keying (BPSK), message A comprises two symbols. In Experiments 2 \& 3, which utilize Quadrature Shift Keying (QPSK), Message A comprises four symbols. In all experiments, Message B was a binary sequence. The lengths of messages A and B need not be equal, and we refer to length of A message as $length_A$ and length of B message as $length_B$ ($length_B$ is typically shorter than $length_A$). In all experiments we assume a sample rate of 1 Hz and oversampling of 1 with respect to A, such that $length_A$ is equal to the number of IQ samples in the signal.

\subsection{Model Architecture}

\begin{figure*}
    \centering
     \begin{subfigure}{\linewidth}
         \includegraphics[width=\linewidth]{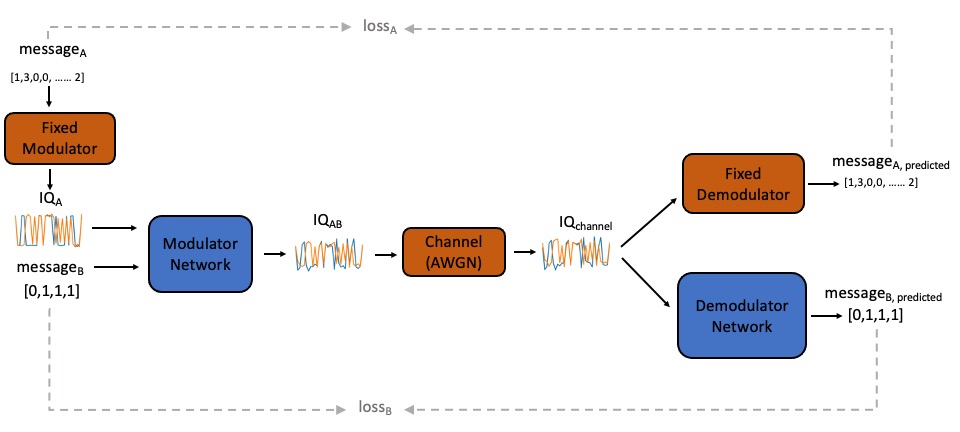}
         \caption{Model Overview.}
         \label{fig:model-overview}
     \end{subfigure}
     \begin{subfigure}{.4\linewidth}
         \centering
         \includegraphics[width=.4\linewidth]{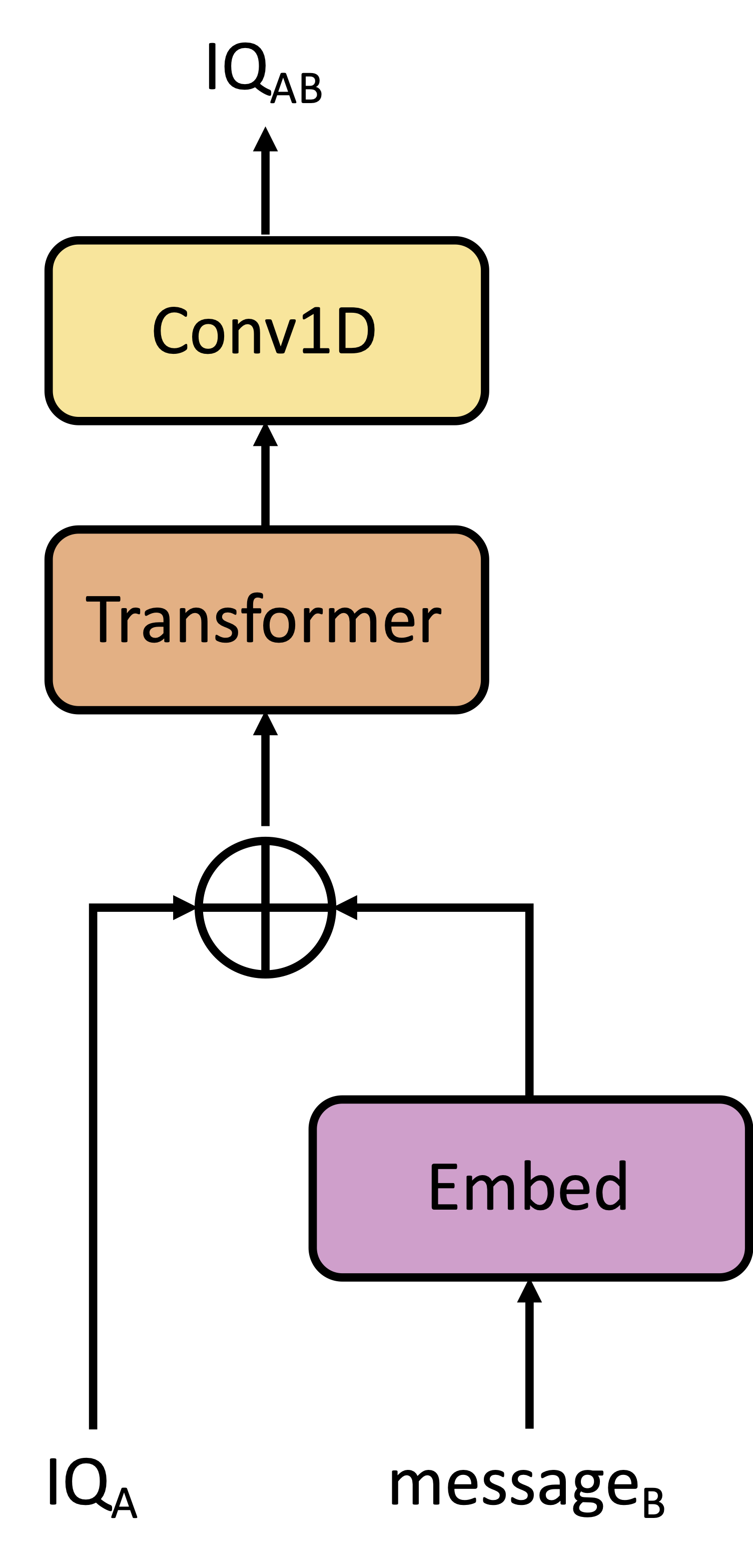}
         \caption{Modulator Network Architecture.}
         \label{fig:mod-net}
     \end{subfigure} %
     \begin{subfigure}{.375\linewidth}
         \centering
         \includegraphics[width=.375\linewidth]{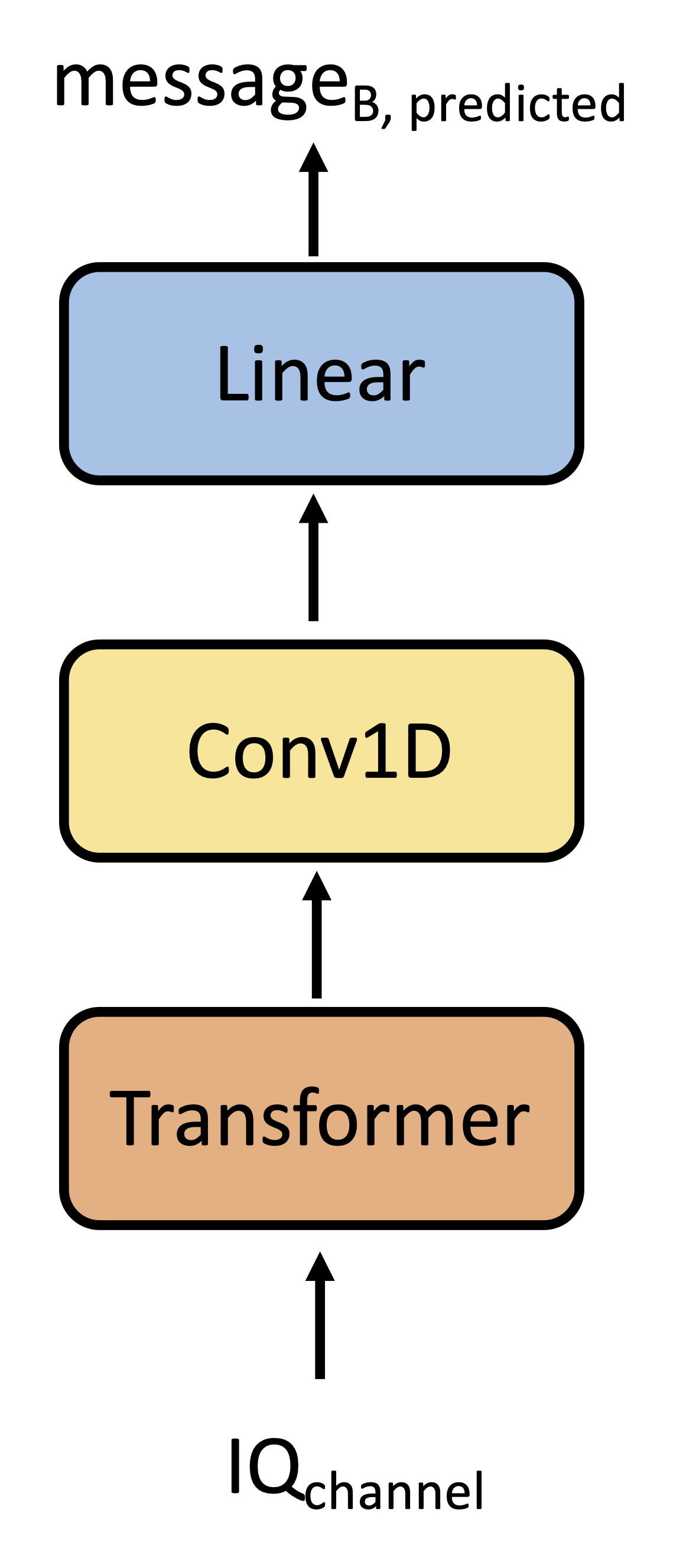}
         \caption{Demodulator Network Architecture.}
         \label{fig:demod-net}
     \end{subfigure}
   \caption{(\ref{fig:model-overview}) High-level model overview. Blue boxes represent learnable, deep neural modules: the Modulator and Demodulator networks. These networks are jointly trained to encode and decode bits from Message B, without disrupting the fixed demodulation. Orange boxes represent fixed modules. (i) The channel module consists of Additive White Gaussian Noise (AWGN), (ii) the fixed modulator converts bits from Message A into an RF signal ($IQ_A$) according to some predefined protocol (e.g., BPSK or QPSK), and (iii) the fixed demodulator processes the learned modulation ($IQ_{channel}$) as if it were a typical BPSK/QPSK signal, and knows nothing about the learned modulation. Grey arrows represent the two main loss terms used in training, for reconstructing Messages A and B. (\ref{fig:mod-net}) Modulator Network architecture. The network receives $IQ_A$ and Message B as inputs, and outputs an IQ signal ($IQ_{AB}$) to be sent over the air. (\ref{fig:demod-net}) Demodulator Network architecture. This network receives $IQ_{channel}$ as input and outputs logits encoding predictions for bits in Message B.}
   \label{fig:model-diagram}
\end{figure*}

As shown in Figure~\ref{fig:model-overview}, our model includes two transformer-based DNNs --- the Modulator and Demodulator networks. These networks are jointly trained to modulate and demodulate extra information from message B without degrading the original signal carrying message A. The model also includes fixed modules for modulation and demodulation of message A, as well as a channel model that simulates Additive White Guassian Noise (AWGN).

Message A is first modulated with a standard RF protocol, such as BPSK or QPSK\footnote{The experiments presented in this paper use BPSK and QPSK as fixed modulations, but in principle this could be substituted for any arbitrary modulation protocol.}. This yields a signal --- an IQ sequence of dimensionality (2, $length_{A}$) --- which we denote $IQ_{A}$. The Modulator Network receives $IQ_{A}$ and message B as inputs, and outputs $IQ_{AB}$, an IQ signal encoding information from both messages. $IQ_{AB}$ is then passed through the channel model, which applies AWGN according to a specified signal-to-noise-ratio (SNR), producing $IQ_{channel}$, a noised signal representing what would be received over-the-air.

The received signal is then separately demodulated by a fixed module, which uses standard demodulation (either BPSK or QPSK) to recover Message A, and the Demodulator Network, which predicts bits in Message B. The discrepancy between ground-truth and predicted symbols in messages A and B serve as two loss terms for training our models, as described in Section~\ref{section:methods/training}. Note that the fixed demodulator is completely naive to the learned modulation; it processes the transmitted signal as if it were a typical BPSK or QPSK signal. Therefore, in order to achieve high accuracy with respect to message A, the Modulator Network must not interfere with the fixed modulation.

Figures ~\ref{fig:mod-net} and ~\ref{fig:demod-net} present a more detailed view of the architectures of the neural network modules. In the Modulator Network, bits from Message B are embedded to a latent space of dimension $\frac{length_A}{length_B}$. These embeddings are then flattened (to match the dimensionality of $IQ_A$), and stacked with $IQ_A$, yielding a 3-dimensional vector containing signal A's I and Q components, as well as a sequence of embeddings representing message B. This vector is fed through a transformer model; we use the Performer implementation \cite{choromanski2020rethinking}, because of increased runtime efficiency. The transformer model consisted of three layers, three self-attention heads, and a dimensionality of three. All other hyper-parameters were set to the defaults specified in the \cite{choromanski2020rethinking} pyTorch package. (Readers are referred to \cite{vaswani2017attention} for explanations of transformer hyper-parameters.) The transformer model outputs a vector of dimensionality (3, $length_A$), which is passed through a 1-dimensional convolutional network with ${kernel size}=1$, $stride=1$, $padding=0$, 3 input channels and 2 output channels. The final output is a single IQ signal, $IQ_{AB}$, encoding information for both messages.

The Demodulator Network is essentially the reverse. $IQ_{channel}$ is passed through a transformer model (dimensionality of 2, 4 layers and 2 self-attention heads), and then convolved (with a 1-dimensional convolutional network with 2 input channels, 1 output channel, kernel size of 1, padding of 0 and stride of 1). The output of the CNN is linearly projected to a 128-dimensional latent space, batch normalized, and then passed through a geLU activation function. The latent vector is then passed through a final linear layer projecting it to a dimensionality of $length_B$. We treat each value in this final vector as a binary prediction logit corresponding to each bit in Message B.

\subsection{Training Procedure}
\label{section:methods/training}

Our models were jointly trained to minimize two loss terms: $loss_A$ and $loss_B$. For $loss_A$ we took the binary cross-entropy (BCE) loss of each IQ sample in $IQ_{channel}$ compared to the original $IQ_A$. For $loss_B$ we took BCE of each prediction logit in the output of the Demodulator Network compared to the bits in the ground-truth Message B. In both cases, prediction logits were passed through a sigmoid function before BCE was computed. These two loss terms were combined into a single loss function that implicitly encouraged the Modulator Network to modulate message B in such a way that it did not degrade original QPSK message. The overall loss is expressed in the equation:

\begin{equation}
{loss} = \alpha\ {loss_A} + (1-\alpha)\ {loss_B}
\label{eq:loss}
\end{equation}

\noindent where $\alpha$ tunes the degree to which $loss_A$ is weighted with respect to $loss_B$. 

Through preliminary experimentation, we found it was best to initialize $\alpha=1$ at the beginning of training (keeping it fixed at 1 for first three epochs), and then gradually decrease it over subsequent epochs (at a rate of 0.01 per epoch) until it reached $\alpha = 0.5$. This encouraged the model to first minimize $loss_A$ --- which should be trivial, since the Modulator Network is given the ground truth QPSK IQ values for message A, and can in principle learn to ignore message B --- and then gradually learn to include information from message B without degrading the original IQ sequence. We also experimented with various auxiliary loss terms for constraining various properties of the generated signals, as described in subsequent sections.

A dataset consisting of 16,384 examples was synthesized. Each example consisted of a tuple of (message A, $IQ_A$, message B). 80\% of these examples were used for training, and the remaining 20\% were held out as a test set. Unless otherwise reported, the batch size was 64, and SNR was varied across all examples within each batch by sampling over a uniform distribution ranging from 5–15 dB. The AdaBelief optimizer was used with a learning rate of 0.01 \cite{zhuang2020adabelief}. Models were trained for 128 epochs, unless otherwise specified.

\section{Results and Discussion}
\label{sec:results}

\subsection{Experiment 1: Proof of concept (BPSK)}
\label{sec:results/bpsk}


Before evaluating our methods on QPSK, we first considered a problem where message A is encoded with BPSK, where $length_A=length_B=32$. Using the training procedure specified above, we successfully trained a model that sends learned bits with high accuracy, and without degrading the original BPSK-modulated message (9.57e-6 Bit Error Rate (BER) for A message, and 0 BER for B message at 14 dB SNR). 

We evaluated the performance gain of our methods by contrasting the channel capacity of a vanilla BPSK-channel carrying just message A, with the capacity of our learned signals carrying messages A and B. Capacity was computed empirically, by multiplying bitrate (in this case, 1 bit/sec for both A and B) times the bitwise accuracy observed at a given SNR.  This was done separately for messages A and B. 

\begin{figure}[h]
    \centering
    \includegraphics[width=.5\linewidth,]{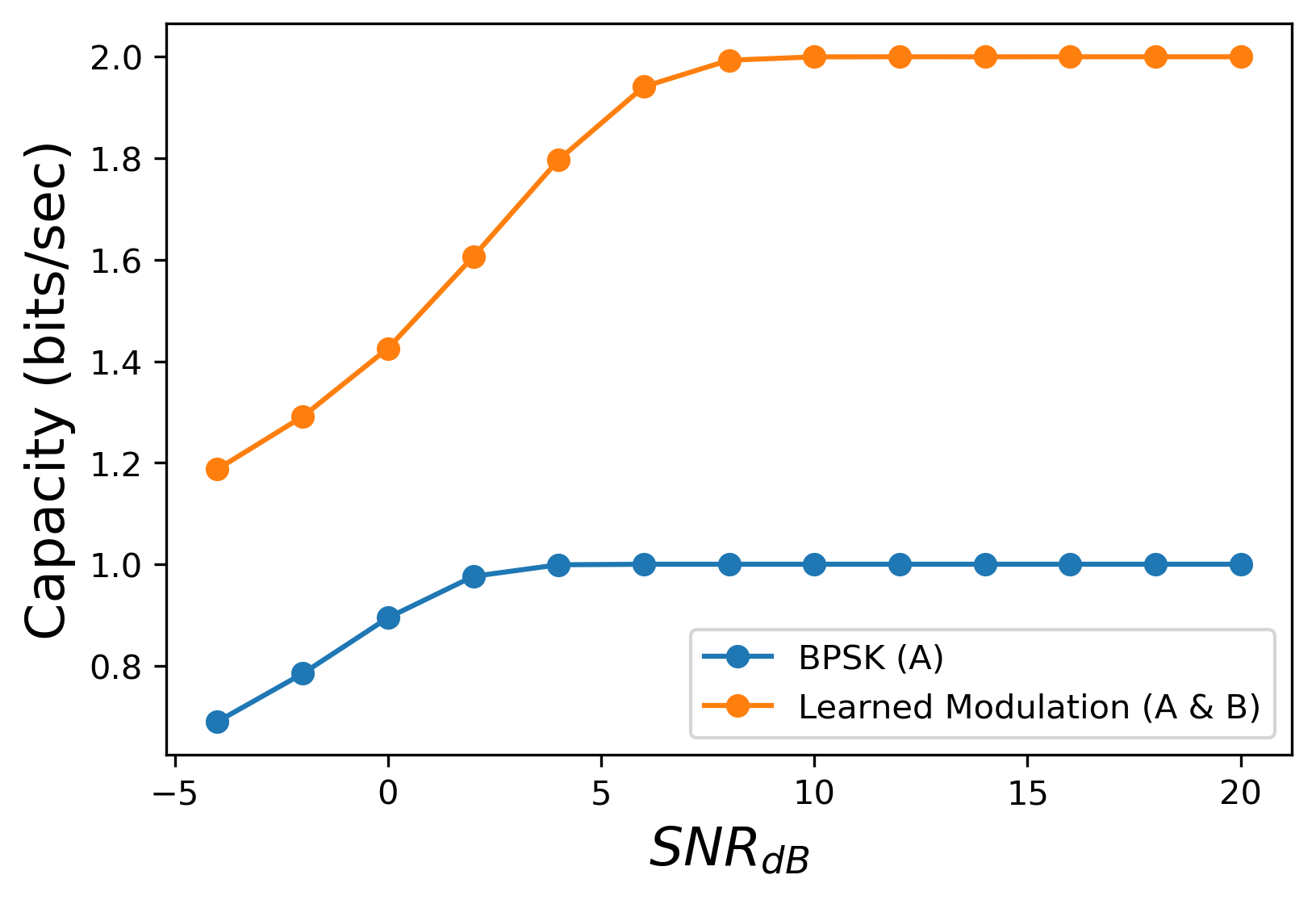}
   \caption{Spectral filling increases capacity of a fixed-bandwidth channel.  The x-axis represents SNR applied to examples in test set. The y-axis represents Capacity, empirically determined by multiplying bitrate times the observed bitwise accuracy at a given SNR. Capacity of original BPSK signal carrying message A is shown in blue, and the learned signal carrying messages A and B is shown in yellow.}
   \label{fig:bpsk-capacity}
\end{figure}

Results are depicted in Figure~\ref{fig:bpsk-capacity}. The blue curve depicts capacity for just message A – this would be channel capacity if only message A was being sent with BPSK. The yellow curve depicts capacity of the learned modulation, sending both A and B. The yellow curve is consistently higher, indicating that our methods effectively increased the capacity of a vanilla BPSK channel across a range of SNRs. More specifically, the capacity of the learned modulation is twice that of the original BPSK signal. This is because near-100\% accuracy was achieved at most SNRs, and by setting $length_A=length_B$, we sent twice the number of bits in the same number of IQ samples.

\subsubsection*{Analysis of Learned Signals}

\begin{figure}[h]
    \centering
    \includegraphics[width=\columnwidth,]{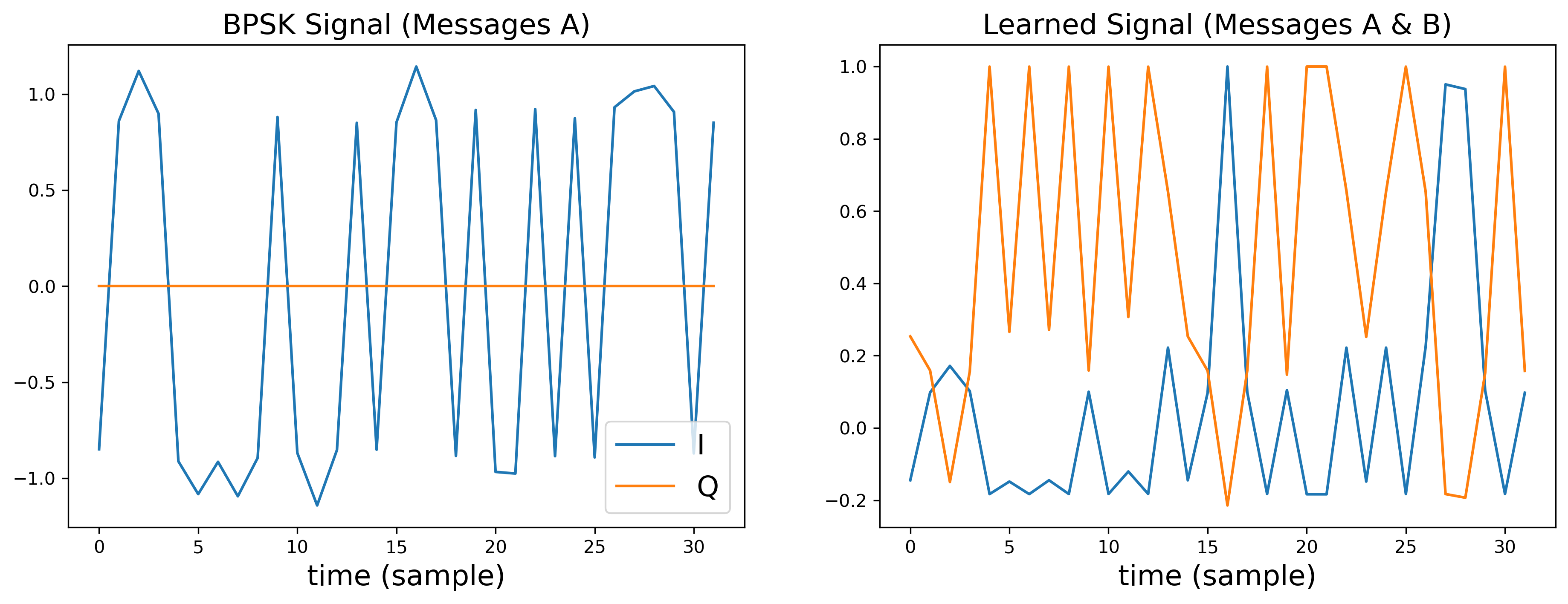}
   \caption{Signals for fixed (BPSK, left) and learned modulations (right) for an arbitrary example. BPSK encodes all information on the I component (blue), leaving the Q component (yellow) at full capacity. The neural network model learns to exploit this, by encoding information for Message B on the Q component.}
   \label{fig:iq-bpsk}
\end{figure}

How did the Modulator Network learn to increase channel capacity without degrading accuracy of the original BPSK signal? To answer this question we compared signals produced by the Modulator Network with the original BPSK signals. Figure~\ref{fig:iq-bpsk} depicts signals corresponding to an arbitrary example in our dataset. The left plot shows the BPSK signal carrying message A, while the right plot shows the learned signal carrying messages A and B. As can be seen, BPSK encodes all information on the I component (blue), leaving the full capacity of orthogonal Q component (yellow) unused. Conversely, the Q component in the learned modulation is highly variable, suggesting that the Modulator Network is utilizing it to encode information from message B. Because these components are orthogonal, the learned modulation is thus able to carry extra information without interfering with message A.

This being said, while the I component has the same essential structure in the BPSK and learned signals, it is scaled in unusual ways in the learned signal. The Q component of the learned signal also looks dissimilar from a typical PSK modulation, in that it is not symmetric around zero. Thus, while the learned signal successfully transmits information from message B without degrading BPSK fidelity, it generates a strange looking signal which could be detected as anomalous by third-party listeners. In the following experiments, we show that by introducing an auxilliary loss term in training, we can constrain learned modulations to highly resemble the fixed modulations.

\subsection{Experiments 2 \& 3: QPSK}

After evaluating our methodology with BPSK modulations, we next turned to a more challenging problem: learning to send extra information over a QPSK-modulated fixed message. This problem is significantly more challenging than the BPSK version, because QPSK encodes Message A on both the I and Q components and thus there is no ``empty'' component for the learned modulation to exploit. In the following subsections we present results from two experiments, which used different auxiliary losses for constraining the structure of the learned signal. In both cases $length_{A} = 1024$ and $length_{B} = 4$. In so doing, we intentionally ``oversampled'' bits from message B.

\subsubsection*{Experiment 2: Constraining learned signals in time-domain (QPSK)}
\label{subsection:mse-aux-loss}

In this experiment we added an auxiliary loss term to explicitly encourage the model to generate signals resembling the original QPSK signal ($IQ_A$). We used mean-squared error (MSE) on the learned IQ sequence ($signal_{combined}$), with respect to the original QPSK signal ($IQ_A$). This loss term is denoted $loss_{MSE}$, and it was incorporated into the overall loss function as defined by the equation:

\begin{equation}
{loss} = \frac{\alpha}{2}\ {loss_A} + (1-\alpha)\ {loss_B} + \frac{\alpha}{2}\ {loss_{MSE}}
\label{eq:mse-loss}
\end{equation}

\noindent This closely resembles Equation~\ref{eq:loss}, except that the weight of $\alpha$ is equally distributed across $loss_A$ and $loss_{MSE}$. This was done because these two loss terms are complementary – constraining $signal_{combined}$ to match $IQ_A$ (via $loss_{MSE}$) necessarily makes it easier for a QPSK demodulator to recover Message A by processing $signal_{combined}$ as if it were a typical QPSK signal. In this sense a high value of $\alpha$ still biases training to optimize for Message A, and low or intermediate $\alpha$ values reward successfully transmitting and demodulating Message B.

\paragraph*{Model Performance}

\begin{figure}
\centering
    \includegraphics[width=.5\columnwidth,]{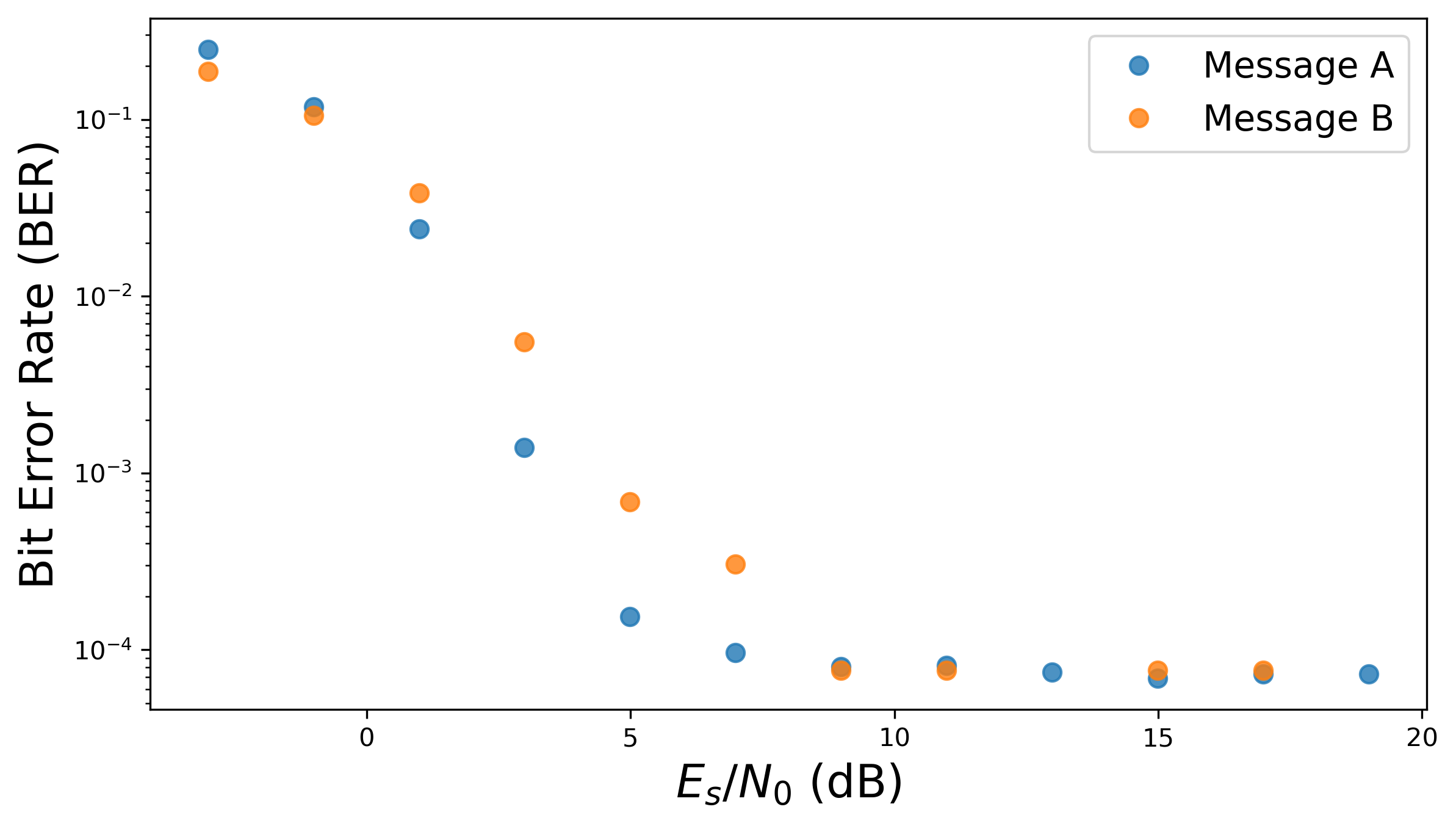}
   \caption{High model accuracy across a range of noise levels. y-axis represents empirically determined Bit Error Rate (BER) for bits from Message A (blue points) and Message B (yellow points). x-axis represents noise level at which our AWGN channel was simulated, expressed in $E_s/N_0$ (energy per symbol to noise power spectral density ratio), a normalized SNR measure. Missing yellow points (at $E_s/N_0~=13$ and $E_s/N_0=18$) are instances where 100\% accuracy was achieved for Message B. Both messages are consistently transmitted and demodulated with high fidelity over a range of noise levels.}
   \label{fig:ber-curve}
\end{figure}

The best model from this training run was evaluated on a held-out test set over a range of SNRs. At each SNR, we passed every example in the test set through the model, and independently evaluated BER of messages A and B. Results are depicted in Figure~\ref{fig:ber-curve}. The x-axis represents noise level at which our AWGN channel was simulated, expressed in $E_s/N_0$ (energy per symbol to noise power spectral density ratio), a normalized SNR measure. The y-axis represents empirically determined BER at each noise-level. Blue points represent BER with respect to message A, and the yellow points represent BER with respect to message B. As expected, BER decreases with more favorable noise levels, until it plateaus at an $E_s/N_0$ of about 8 dB. Most importantly, the model achieves an acceptably low BER for both messages, and this is robust across a range of SNRs.\footnote{It is also worth noting that these BER values can be further enhanced with forward-error correction strategies \cite{mackay1999good, nafaa2008forward}, which would be straightforward to integrate with our model.}

In response to our primary research question, this demonstrates the ability to successfully learn a modulation that can transmit extra information (Message B) in the same channel as a fixed-modulation signal without degrading the original signal. Whereas the previous subsection demonstrated the feasibility of our methods with BPSK-modulated signals, here we observe success with respect to QPSK, which is substantially more difficult because information from Message A is spread across both IQ components.

\begin{figure*}[ht!]
     \begin{subfigure}{.5\textwidth}
         \centering
         \includegraphics[width=\textwidth]{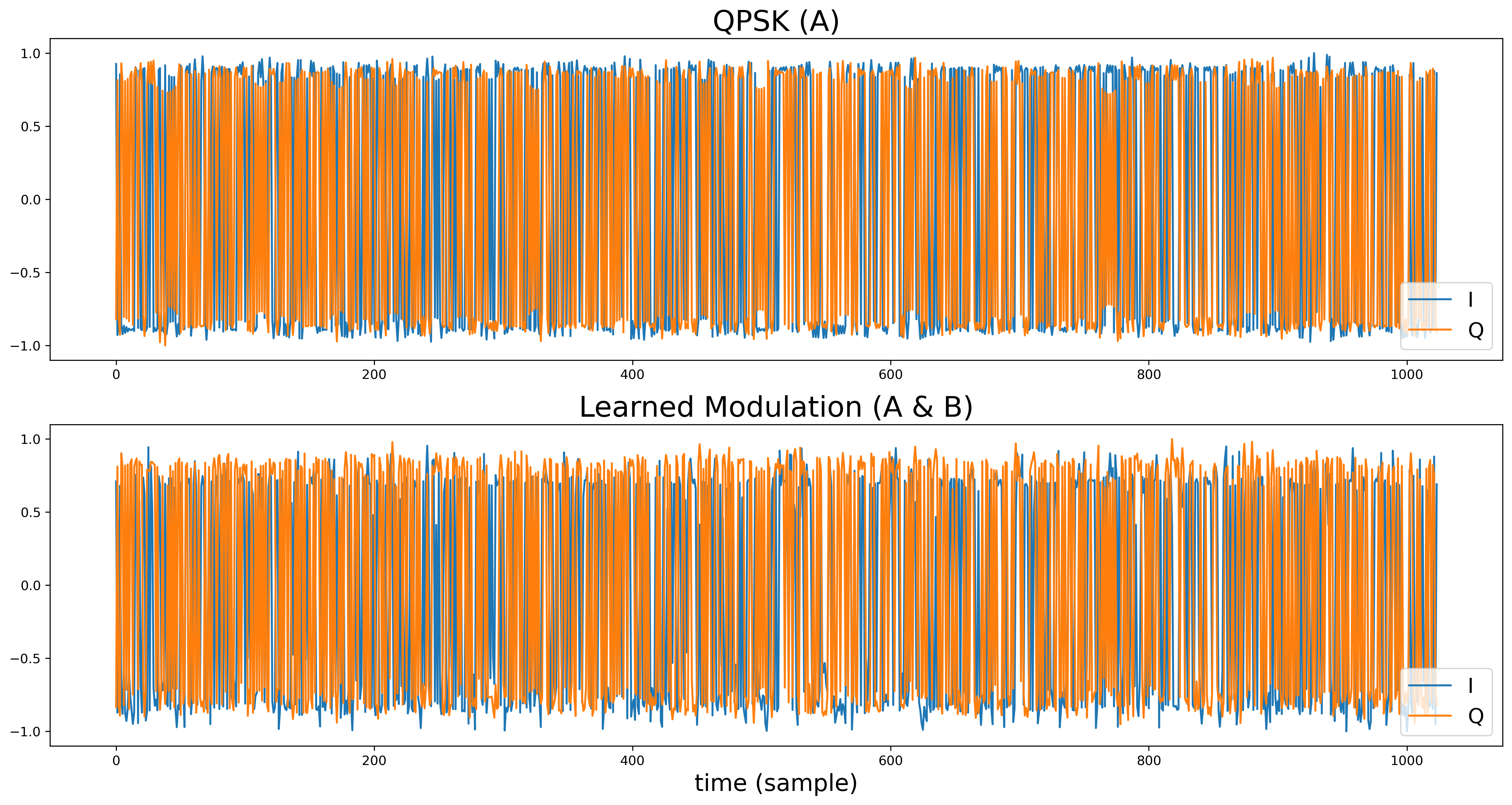}
         \caption{Time-domain signals.}
         \label{fig:qpsk-example-time}
     \end{subfigure}
     \begin{subfigure}{.5\textwidth}
         \centering
         \includegraphics[width=\textwidth]{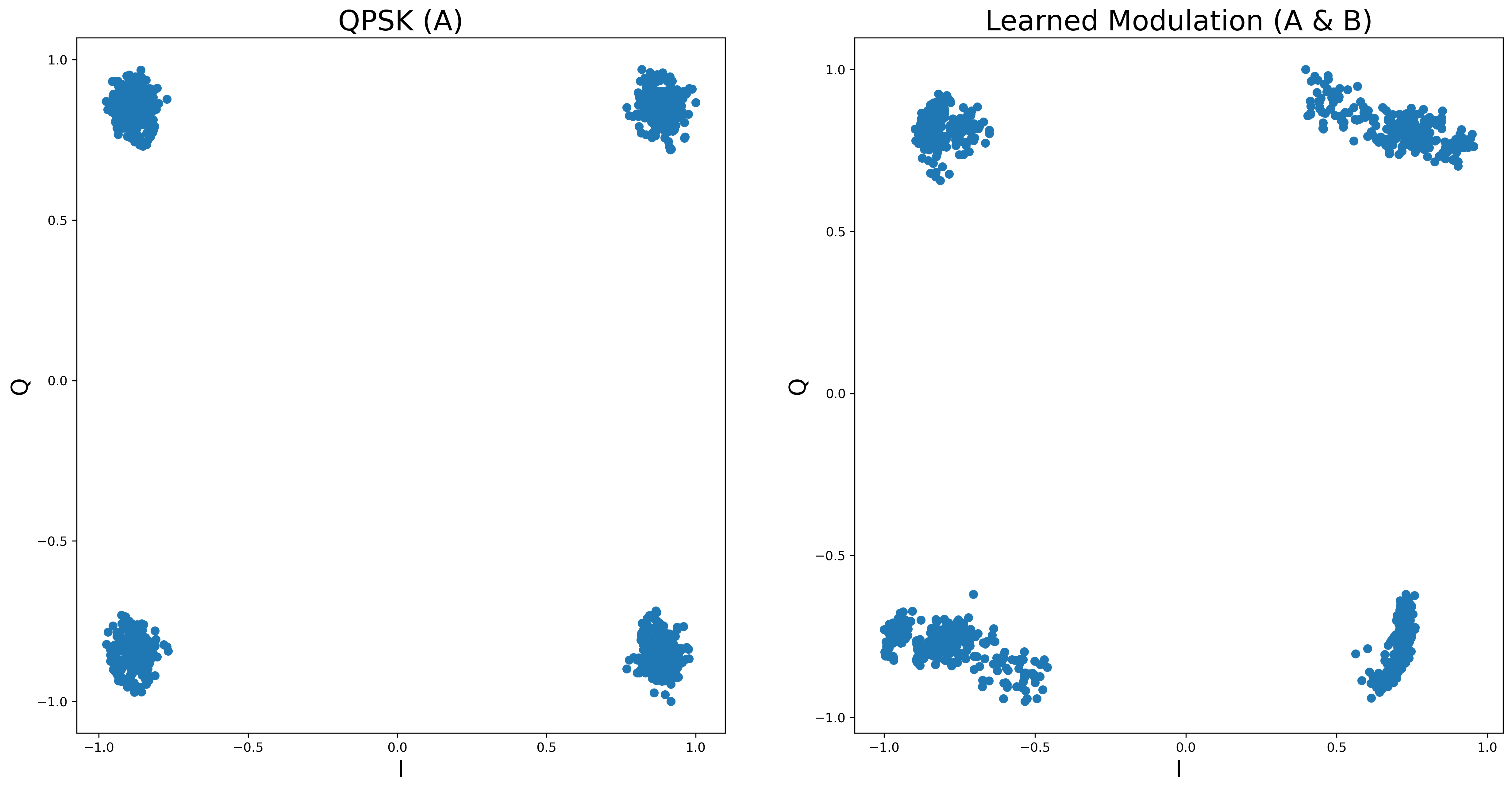}
         \caption{Constellation plots.}
         \label{fig:qpsk-example-constellation}
     \end{subfigure}
     \caption{Fixed (QPSK-modulated) and learned signals for an arbitrary example. The learned signal carries Message A and B, whereas the QPSK signal only carries Message A. There is a high resemblance between the two signals, demonstrating that it is possible to learn a modulation for send additional information around an existing signal without substantively altering the original modulation.}
     \label{fig:qpsk-example-iq}
\end{figure*}

Next we turn to our secondary research question: can we constrain the structure of learned signals? In this experiment we were interested in constraining the learned signal to match the original $signal_A$. To get a sense of this, we visualized examples of learned signals generated by our best-performing model, and compared them to the original QPSK signals. Figure~\ref{fig:qpsk-example-iq} depicts an arbitrary example, with time-domain plots in Figure~\ref{fig:qpsk-example-time} and constellation plots in Figure~\ref{fig:qpsk-example-constellation}. From both vantage points, the learned signal closely resembles the original QPSK signal. Thus, not only did our model successfully learn to transmit information from both messages, it did so in such a way that the learned signals were nearly identical to original modulations. This has important implications for our methods in real world RF applications – we can learn to transmit extra information in ``hidden'' messages, such that the generated signals look nearly identical to typical QPSK signals from the perspective of a third-party.

\subsubsection*{Experiment 3: Constraining Constellation Plots of Learned Signals (QPSK)}
\label{sec:exp-3}

We also experimented with different methods for constraining the constellation plots of learned signals. As in Experiment 2, this could be done with an aim towards producing a learned signal that maximally resembles the fixed modulation, but we show that these techniques can also be used for arbitrary signal shapes.

To learn a particular shape, we add an auxiliary term to the loss function that encourages the distribution of values in the learned signal to match a target distribution.  Given a sample of $m$ points from a target distribution and a learned signal, we define the $n \times m$ distance matrix $M$ as:  $M_{ij}=MSE(s_i,q_j )$ where each $s_i$ is a single value sampled from the learned signal, and each $q_i$ is from a sample of the target distribution . The auxiliary loss is then:

\begin{equation}
	loss_{shape} = \frac{1}{n} \sum_{i=0}^{n} min_j (M_{ij}) + \frac{1}{m} \sum_{j=0}^{m} min_i (M_{ij})
	\label{eq:loss-shape-term}
\end{equation}

\noindent This first sum encourages each learned signal value to be near a point in the target distribution sample.  The second term ensures that the learned signal shape takes on the entire target structure. For example, in the case of a multimodal distribution, without the second loss term, the shape loss could be minimized if all the learned signal points cluster on one of the modes. Notably, this loss function does not require a closed-form density function for the target distribution so the learned signal can resemble any shape compatible with QPSK or another established communication protocol.  The complete loss equation becomes: 

\begin{equation}
	loss = \alpha\ loss_A + (1 - \alpha)\ loss_B + \beta\ loss_{shape}
	\label{eq:loss-shape}
\end{equation}

\begin{figure*}[h]
\centering
    \includegraphics[width=\linewidth]{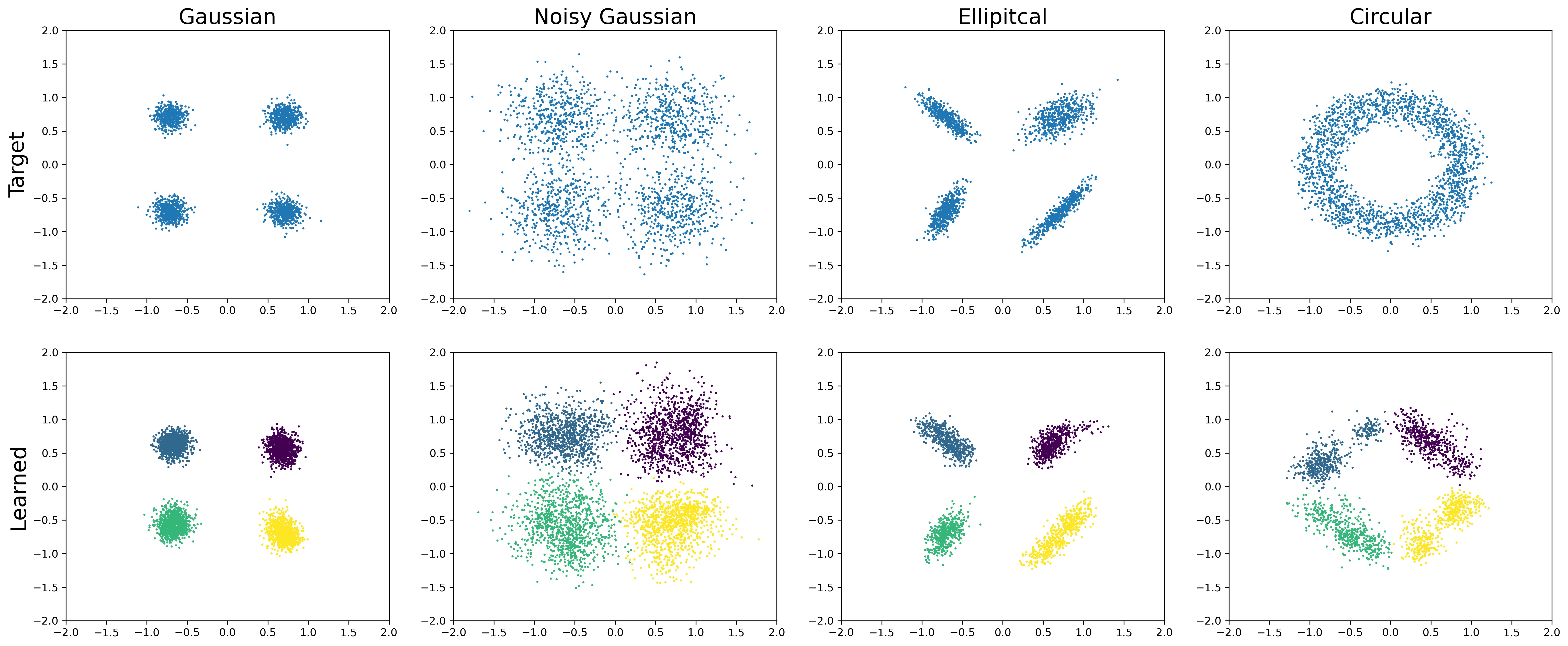}
   \caption{Constraining learned modulation to arbitrary spectral shapes. Top row presents constellation plots of target distributions, and bottom row presents constellation plots of learned signals, color-coded by the ground-truth QPSK symbol encoded by each IQ sample. Using an auxiliary loss term, we were able to constrain generated signals to conform to arbitrary spectral shapes, while still retaining high fidelity with respect to both messages.}
   \label{fig:spectral-filling-shapes}
\end{figure*}

We have tested this with several shapes and depict the results in Figure~\ref{fig:spectral-filling-shapes}. From left to right, the first two shapes were trained to resemble a QPSK signal at different noise levels. We trained at a fixed SNR of 10 dB for 50 epochs.  For computational efficiency, the distance matrix was calculated using 2500 random values from the learned signal and 2500 random values from the target distribution. BER at SNR = 10 dB for the A message was 1.49e-7 and 5.19e-3 for the less noisy and noisy targets, respectively.  The BER for the B message was zero for both models.

The other two shapes demonstrate the flexibility of this method. For these two shapes, we trained with SNR fixed at 10 dB for 200 epochs using the sampling adjustment described in the previous paragraph. For the elliptical distribution the BER for the A message was 1.71e-5 and for the B message was 0.  For the circular distribution the BER for the A message was 2.4e-3 and 7.62e-5 for the B message. Thus, these methods allow us to constrain generated signals to conform to arbitrary spectral shapes, while still retaining high fidelity with respect to both messages.

\section{Conclusion}
\label{sec:conclusion}

We have demonstrated the ability to use deep, transformer-based neural networks for ``spectral filling.'' Given an original message (Message A), encoded with some pre-defined modulation protocol (e.g., BPSK/QPSK), these networks can learn to augment and reconstruct the IQ sequence, such that it carries an additional message (Message B) without degrading the original signal. This has promising implications for congested IoT applications, as it establishes a methodology for increasing the capacity of existing fixed-bandwidth RF channels without costly human-engineered protocols, and without disrupting existing communications protocols. This last point is crucial, because a major challenge in leveraging generative deep learning for RF applications is how to deploy these technologies without disrupting pre-established RF environments.

We have further demonstrated that with the help of auxiliary loss terms, it is possible to constrain learned signals to closely resemble the original signals, or to match arbitrary spectral shapes, while still transmitting information from both messages at high fidelity.  The fact that extra information can be sent without significantly altering the original signal means this technique can be used in sensitive contexts, to send additional \textit{in cognito} messages, undetectable to third-party listeners.


\section{References}

\bibliographystyle{plain}
\bibliography{main}

\end{document}